\documentstyle[amssymbols,fullpage]{article}

\newcommand{\beq}{\begin{equation}}
\newcommand{\eeq}{\end{equation}}
\newcommand{\bea}{\begin{eqnarray}}
\newcommand{\bean}{\begin{eqnarray*}}
\newcommand{\eea}{\end{eqnarray}}
\newcommand{\eean}{\end{eqnarray*}}
\newcommand{\demi}{\frac{1}{2}}

\newcommand{\sumi}{\sum_{i=1}^N}
\newcommand{\sumij}{\sum_{_{\ i \neq j}^{i,j=1}}^N}
\newcommand{\la}{\lambda}
\newcommand{\si}{\sigma}

\newcommand{\Si}{\Sigma_\alpha}

\title{Exact Yangian Symmetry \\ in the classical Euler-Calogero-Moser Model}
\author{E. Billey $^*$ \and J. Avan $^*$ \and O. Babelon
\thanks{L.P.T.H.E. Universit\'e Paris VI (CNRS UA 280), Box 126, Tour 16,
$1^{er}$ \'etage, 4 place Jussieu, F-75252 PARIS CEDEX 05}}
\date{January 1994}

\begin{document}

\def\N{\Bbb N}
\def\C{\Bbb C}

\begin{titlepage}
\renewcommand{\thepage}{}
\maketitle
\vspace{2cm}
\begin{abstract}
We compute the $r$-matrix for the elliptic Euler-Calogero-Moser model. In
the trigonometric limit we show that the model possesses an exact Yangian
symmetry.
\end{abstract}

\vfill

PAR LPTHE 94-03

\end{titlepage}
\renewcommand{\thepage}{\arabic{page}}

\section{Introduction}
The Euler-Calogero-Moser model was defined in \cite{GH,W}. In \cite{BAB}
we considered the rational case and we derived the $r$-matrix. In this paper
we are interested in its trigonometric and elliptic generalizations. In the
elliptic case we compute the $r$-matrix and show that the usual elliptic
Calogero-Moser model and its $r$-matrix are obtained by Hamiltonian
reduction. In the trigonometric case we show that the current algebra
symmetry exhibited by Gibbons and Hermsen \cite{GH} in the rational case, is
deformed into a Yangian symmetry algebra.

We consider a system of $N$ particles on a line with pairwise interactions.
The degrees of freedom consist of the positions and momenta
$(p_i,q_i)_{i=1 \cdots N}$ and of antisymmetric additional variables
$(f_{ij}=-f_{ji})_{i,j=1 \cdots N}$, with the Poisson brackets
\bea
  \label{pb1}
  \{ p_i, q_j \} & = & \delta_{ij} \\
  \label{pb2}
  \{ f_{ij}, f_{kl} \} & = & \demi \ ( \delta_{il} \ f_{jk} +
                                 \delta_{ki} \ f_{lj} +
                                 \delta_{jk} \ f_{il} +
                                 \delta_{lj} \ f_{ki}  ) .
\eea
The Poisson brackets of the $f_{ij}$ just reproduce the $O(N)$ Lie algebra.
The Hamiltonian will be taken of the form
\beq
  \label{Ham}
  H = \demi \sumi p_i^2 - \demi \sumij f_{ij} \ f_{ji} \ V(q_{ij}) ,
  \ \ \ \ \  q_{ij}=q_i-q_j
\eeq
with an even potential $ V(-x)=V(x).$

The equations of motion are easily derived:
\bean
  \dot{q}_i & = & p_i \\
  \dot{p}_i & = & \sum_{_{j \neq i}^{j=1}}^N  f_{ij} \ f_{ji}  \ V'(q_{ij}) \\
  \dot{f}_{ij} & = & \sum_{_{k \neq i,j}^{\ k=1}}^N  \ f_{ik} \ f_{jk} \
        \left [ V(q_{ik}) -  V(q_{jk}) \right ].
\eean
Such a system admits a Lax representation only for specific potentials.
Indeed writing the following ansatz for the Lax pair
\bea
  L(\la) & = & \sumi p_i \ e_{ii} + \sumij l(q_{ij},\la) \ f_{ij} \ e_{ij} \\
  M(\la) & = & \sumij m(q_{ij},\la) \ f_{ij} \ e_{ij}
\eea
where $e_{ij}$ is the $N \times N$ matrix $(e_{ij})_{kl}= \delta_{ik} \
\delta_{jl}$ and $\la \in \C$ is the spectral parameter, we find that the
equations of motion can be written in the Lax form
\beq
  \label{Lax}
  \dot{L}(\la)=[M(\la),L(\la)]
\eeq
if and only if the following equalities are satisfied:
\bea
  m(x,\la) & = & - \frac{\partial}{\partial x}l(x,\la) = -l'(x,\la)  \\
  \label{Cf}
  l'(x,\la) \ l(y,\la) - l'(y,\la) \ l(x,\la) & = & l(x+y,\la)
  \ [ V(x) - V(y) ] \\
  l(x) & \sim & - \frac{1}{x} \ \ \mbox{when} \ \ x \to 0.
\eea
Eq.(\ref{Cf}) is the famous functional equation of Calogero. Its general
solution is \cite{C1,C2}
\beq
  l(x,\la) = - \frac{\si(x+\la)}{\si(x)\ \si(\la)}, \ \ \ \ \
  V(x) = \wp(x)
\eeq
where $\si$ and $\wp$ are Weierstrass elliptic functions, the relevant
properties of which are recalled in the appendix. The elliptic $O(N)$
Euler-Calogero-Moser model is precisely defined by eq.(\ref{Ham}) with
$ V(x)=\wp(x)$ together with the Poisson brackets (\ref{pb1},\ref{pb2}).

\section{The $r$-matrix}
 From eq.(\ref{Lax}) it follows that $ trL^n(\la) $ is a set of conserved
quantities. In particular
$$ trL(\la) = \sumi p_i , \ \ \ \ \ trL^2(\la) = 2 H + \wp(\la). $$
The involution property of these quantities  $trL^n(\la)$
will follow from the existence of an $r$-matrix which  we now calculate
\cite{S1,BV}. Introducing the notations $L_1(\la)=L(\la) \otimes 1$ and
$L_2(\la)= 1 \otimes L(\la)$ we show that the Poisson brackets of the Lax
matrix elements can be recast as
\beq
  \{ L_1(\la),L_2(\mu) \} = [r_{12}(\la,\mu),L_1(\la)]
                           -[r_{21}(\mu,\la),L_2(\mu)].
\eeq
Following \cite{BAB} we assume that $r$ is of the form
$$  r_{12}(\la,\mu) =  a(\la,\mu) \sumi e_{ii} \otimes e_{ii}
                    + \sumij b_{ij}(\la,\mu) e_{ij} \otimes e_{ji}
                    + \sumij c_{ij}(\la,\mu) e_{ij} \otimes e_{ij}. $$
Requiring that $r_{12}(\la,\mu)$ be independent of the $p_i$ variables we
obtain
\bea
  b_{ij}(\la,\mu) & = & - b_{ji}(\mu,\la) \\
  c_{ij}(\la,\mu) & = & c_{ij}(\mu,\la).
\eea
Moreover assuming that $r_{12}(\la,\mu)$ is independent of the $f_{ij}$
variables yields the following system:
\bea
  \label{1}
  & &a(\la,\mu) \ l(q_{ij},\la)  - b_{ij}(\la,\mu) \ l(q_{ij},\mu)
  + c_{ij}(\la,\mu) \ l(q_{ji},\mu) = - l'(q_{ij},\la) \\
  \label{2}
  & & b_{ij}(\la,\mu) \ l(q_{jk},\la) - b_{ik}(\la,\mu) \ l(q_{jk},\mu) =
  - \demi \ l(q_{ik},\la) \ l(q_{ji},\mu) \\
  \label{3}
  & & c_{ij}(\la,\mu) \ l(q_{jk},\la) + c_{ik}(\la,\mu) \ l(q_{kj},\mu) =
  \demi \ l(q_{ik},\la) \ l(q_{ij},\mu) \\
  \label{4}
  & & c_{ij}(\la,\mu) \ l(q_{ki},\la) + c_{kj}(\la,\mu) \ l(q_{ik},\mu) =
  \demi \ l(q_{kj},\la) \ l(q_{ij},\mu).
\eea
A solution to these equations is
\bean
  a(\la,\mu) & = & - \demi \ [ \zeta(\la+\mu)+\zeta(\la-\mu) ] \\
  b_{ij}(\la,\mu) & = & \demi \ l(q_{ij},\la-\mu) \\
  c_{ij}(\la,\mu) & = & \demi \ l(q_{ij},\la+\mu).
\eean
Indeed substituting the preceding expressions in eq.(\ref{2},\ref{3},\ref{4})
leads to the same relation:
$$ l(q_{ij},\la-\mu) \ l(q_{jk},\la) + l(q_{ki},\mu-\la) \ l(q_{jk},\mu)
                                    + l(q_{ik},\la) \ l(q_{ji},\mu) = 0 $$
which upon setting $ x = \demi \ (\la+q_{ij}), \ y = \demi \ (2\mu-\la+q_{ji})
, \ z =  \demi \ (\la+q_{ji}) $ and $ t = \demi \ (-\la-q_{ki}+q_{kj})$ is
a direct consequence of relation (\ref{a3}).
The expression for $a(\la,\mu)$ is then given by eq.(\ref{1}), and is
simplified using eq.(\ref{a1}) and (\ref{a4}). Finally the $r$-matrix
reads
\bea
   r_{12}(\la,\mu) & = & \demi \ \sumij l(q_{ij},\la-\mu)e_{ij}
                         \otimes e_{ji}
                    + \demi \ \sumij l(q_{ij},\la+\mu)e_{ij}
                         \otimes e_{ij} \nonumber \\
                    & - & \!  \demi \ [ \zeta(\la+\mu)+\zeta(\la-\mu) ]
                         \sumi  e_{ii} \otimes e_{ii}.
\eea

\section{The $sl(N)$ model}\label{not}
The above $O(N)$ model can be obtained from the more general $sl(N)$
model by a mean procedure \cite{Mi,FR,AT}. The $sl(N)$ elliptic Euler-Calogero
Moser model
is defined by the Hamiltonian
\beq
  \label{H}
  H  =  \demi \sumi p_i^2 - \demi \sumij f_{ij} \ f_{ji} \ \wp(q_{ij})
\eeq
and the Poisson brackets
\bea
  \{ p_i, q_j \} & = & \delta_{ij} \\
  \label{Pm}
  \{ f_{ij},f_{kl} \} & = & \delta_{jk} \ f_{il} - \delta_{li} \ f_{kj}.
\eea
For this model we define a Lax matrix as
\beq
  L(\la) = \sumi \left ( p_i - \zeta(\la) f_{ii} \right ) \ e_{ii}
         + \sumij l(q_{ij},\la) \ f_{ij} \ e_{ij}.
\eeq
The Hamiltonian is given by $ H = \demi \int \frac{d\la}{2i \pi \la}
trL^2(\la).$
A direct calculation gives
\bea
  \label{ll}
  \{ L_1(\la),L_2(\mu) \} & = & [r_{12}(\la,\mu),L_1(\la)]
                               -[r_{21}(\mu,\la),L_2(\mu)] \nonumber \\
                 & & \! \! - \sumij l'(q_{ij},\la-\mu) \ (f_{ii}-f_{jj}) \
                           e_{ij} \otimes e_{ji}
\eea
with the beautifully simple $r$-matrix
\beq
  \label{r}
  r_{12}(\la,\mu) = - \zeta(\la-\mu) \sumi e_{ii} \otimes e_{ii}
                    + \sumij l(q_{ij},\la-\mu) \ e_{ij} \otimes e_{ji}.
\eeq
At this point let us make two remarks:
\begin{itemize}
\item Because of the third term in the right member of eq.(\ref{ll}) the
integrals of motion $trL^n(\la)$ are not in involution. However we can
restrict ourselves to the manifolds $ (f_{ii} = \mbox{constant})_
{i=1 \cdots N} $ since $trL^n(\la)$ Poisson-commute with $f_{ii}.$
On these manifolds $trL^n(\la)$ are in involution.
\item The $r$-matrix for the $O(N)$ model eq.(\ref{r}) is immediately seen
to be of the form
$$  r_{12}^{O(N)} = \demi \ ( 1+\si \otimes 1 ) r_{12}^{sl(N)} $$
where $\si$ is the involutive automorphism
$$ \si : \la^n e_{ij} \longmapsto - (-\la)^n e_{ij}. $$
This is typical of a mean construction.
\end{itemize}
In the following we will restrict the $f_{ij}$ to a symplectic leaf of
the Poisson manifold (\ref{Pm}). Introducing vectors
\bean
  (\xi_i)_{i=1 \cdots N} & \ \mbox{with} & \xi_i=(\xi_i^a)_{a=1 \cdots r } \\
  (\eta_i)_{i=1 \cdots N} & \ \mbox{with} & \eta_i=(\eta_i^a)_{a=1 \cdots r }
\eean
with the Poisson brackets
\beq
  \{ \xi_i^a ,\xi_j ^b \} = 0 , \ \ \ \ \ \{ \eta_i^a , \eta_j^b \} = 0 ,
  \ \ \ \ \ \{ \xi_i^a ,\eta_j^b \} = - \delta_{ij} \ \delta_{ab} ,
\eeq
we parametrize the $f_{ij}$ as follows:
\beq
  \label{sm}
  f_{ij}= \langle  \xi_i | \eta_j \rangle = \sum_{a=1}^r \xi_i^a \eta_j^a.
\eeq
The phase space now becomes a true symplectic manifold.

\section{The $r$-matrix of the elliptic Calogero model}
We show here that the $r$-matrix for the elliptic Calogero model
\cite{S1,BS} can be obtained from eq.(\ref{ll}) by a Hamiltonian reduction
procedure \cite{Mi,FR,AT}.

We choose $r=1$ in eq.(\ref{sm}). On the manifold $f_{ij}= \xi_i \ \eta_j$
acts an Abelian Lie group
\beq
  \xi_i \longrightarrow \la_i \ \xi_i , \ \ \
 \eta_i \longrightarrow \la_i^{-1} \eta_i.
\eeq
Remark that the group acts on $L(\la)$ as conjugation by the matrix
$ \mbox{diag}(\la_i)_{i=1 \cdots N}$ and therefore all the Hamiltonians
$trL^n(\la)$ are invariant. Thus one can apply the method of Hamiltonian
reduction.
The infinitesimal generator of this action is
$$ H_{\epsilon} = \sumi \epsilon_i \ f_{ii}, \ \ \ \ \ \la_i=1+\epsilon_i. $$
We fix the momentum by choosing
$$ f_{ii}=\alpha. $$
To compute the reduced Poisson brackets of the Lax matrix, we remark that
the matrix
\bea
  L^{Cal}(\la) & = & g^{-1} L(\la) \ g  \ \ \ \ \ \mbox{with} \
                    g = \mbox{diag}(\xi_i)_{i=1 \cdots N}  \nonumber \\
               & = & \sumi [ p_i - \alpha \zeta(\la) ] \ e_{ii}
                     + \alpha \sumij l(q_{ij,\la}) \ e_{ij}
\eea
is invariant under the isotropy group $G_{\alpha}$ of $\alpha$ (which is
the whole group itself since it is Abelian) and we can compute the Poisson
brackets of its matrix elements directly. We find
\beq
  \label{llcal}
  \{ L_1^{Cal}(\la),L_2^{Cal}(\mu) \} =
                [r_{12}^{Cal}(\la,\mu),L_1^{Cal}(\la)]
               -[r_{21}^{Cal}(\mu,\la),L_2^{Cal}(\mu)]
\eeq
with
$$ r_{12}^{Cal}(\la,\mu) = g_1^{-1} g_2^{-1}
    \left [ \ r_{12}(\la,\mu)  - \{ g_1,L_2(\mu) \} g_1^{-1}
             + \demi [ u_{12},L_2(\mu) ] \right ] g_1 g_2 $$
where $ u_{12} = \{ g_1,g_2 \} g_1^{-1} g_2^{-1} $ is here equal to zero.
Redefining
$$ r_{12}^{Cal}(\la,\mu) \longrightarrow r_{12}^{Cal}(\la,\mu)+
 \left [ \frac{1}{2\alpha} \sumi e_{ii} \otimes e_{ii} , L_2(\mu) \right ] $$
does not change eq.(\ref{llcal}) and yields exactly the $r$-matrix found in
\cite{S2,BS}
\bea
   r_{12}^{Cal}(\la,\mu) & = &
     \sumij l(q_{ij},\la-\mu) \  e_{ij} \otimes e_{ji}
   + \demi \ \sumij l(q_{ij},\mu) \ (e_{ii}+e_{jj}) \otimes e_{ij}
         \nonumber \\
  & &  - [\zeta(\la-\mu) + \zeta(\mu)] \sumi e_{ii} \otimes e_{ii}.
\eea

\section{Yangian symmetry in the trigonometric case}
The parametrization (\ref{sm}) of $f_{ij}$ introduces a $sl(r)$ symmetry
into the theory. The transformation
\bean
  \eta_i^a & \longrightarrow & \sum_{b=1}^r u^{ab} \eta_i^b \\
  \xi_i^a  & \longrightarrow & \sum_{b=1}^r (u^{-1})^{ab} \xi_i^b
\eean
leaves the $f_{ij}$ invariant and therefore also the Hamiltonians.
This symmetry is generated by a set
of conserved currents
\beq
  J_0^{ab} = \sumi \xi_i^b \eta_i^a.
\eeq
It is remarkable that this current was shown, in the rational case
\cite{GH}, to be the first of a hierarchy building a current algebra
commuting with the Hamiltonian --- and more generally with a subset of the
commuting Hamiltonians.

We now extend this result to the trigonometric case, and we will show that
the hierarchy of currents form a Yangian symmetry in this case. Taking the
trigonometric limit ($\omega_1=\infty$ and $\omega_2=i \frac{\pi}{2}$) in
the above formulas, we see that the Lax matrix
can be taken of the form
\beq
  \label{L}
  L(\la)= L_0 - \coth(\la) F
\eeq
with
\beq
  L_0 = \sumi p_i e_{ii} - \sumij \coth(q_{ij}) \ f_{ij} \ e_{ij} , \ \ \ \ \
  F = \sum_{i,j=1}^{N} f_{ij} \ e_{ij}.
\eeq
By a straightforward calculation, or taking the limit of the elliptic case,
we find
\bea
  \label{calg}
  \{ L_1(\la),L_2(\mu) \} & = & [r_{12}^0,L_1(\la)]
                                   -[r_{21}^0,L_2(\mu)] \nonumber \\
             & & \! - \demi \left (1 - \coth(\la) \coth(\mu) \right )
                 \left ( [{\cal C}, F_1]-[{\cal C}, F_2] \right ) \nonumber \\
             & & \! \! - \sumij (f_{ii}-f_{jj}) \frac{1}{\sinh^2(q_{ij})}
                                  e_{ij} \otimes e_{ji}
\eea
where
\beq
r_{12}^0 = - \sumij \coth(q_{ij}) \ e_{ij} \otimes e_{ji}
\eeq
and ${\cal C}$ is the Casimir element of $sl(N)$
\beq
  {\cal C} = \sum_{i,j=1}^N e_{ij} \otimes e_{ji}.
\eeq
In spite of the unusual second term in eq.(\ref{calg}) the quantities
$trL^n(\la)$ are still in involution on the manifolds $ \Si : (f_{ii} =
\alpha)_
{i=1 \cdots N} $. Indeed,
\bean
  \{ trL^n(\la) , trL^m(\mu) \} & = & n \ m \sumij \frac{f_{ii}-f_{jj}}
   {\sinh^2(q_{ij})} \ [L^{n-1}(\la)]_{ij} \ [L^{m-1}(\mu)]_{ji} \\
   & & - \frac{n \ m}{2} (1-\coth(\la) \coth(\mu)) \
    tr_{12} \left ( L_1^{n-1}(\la) L_2^{m-1}(\mu) [ {\cal C},F_1-F_2 ] \right )
\eean
and since $tr_2 \ ( (1 \otimes A) \ {\cal C} ) = A$, we obtain
\bean
  \{ trL^n(\la) , trL^m(\mu) \} & = & n \ m \sumij \frac{f_{ii}-f_{jj}}
   {\sinh(q_{ij})^2} \ [L^{n-1}(\la)]_{ij} \ [L^{m-1}(\mu)]_{ji} \\
    & & - \frac{n \ m}{2} (1-\coth(\la) \coth(\mu)) \
    tr \{ L^{n-1}(\la) [ L^{m-1}(\mu),F]
              - L^{m-1}(\mu) [ L^{n-1}(\la),F ] \}.
\eean
If we notice that
$$ F = - \frac{L(\la)-L(\mu)}{\coth(\la)-\coth(\mu)} $$
we immediately get the involution property.

We consider now the subset $ tr(L^n) = tr(L_0+F)^n$ of commuting Hamiltonians;
notice that $H$ belongs to this subset, since $ H = \demi trL^2 - \alpha \ trL
+ \demi N \alpha^2.$

We introduce the following quantities:
\beq
  J_n^{ab} = tr(L^n F^{ab}) , \ \ \ \ \ a,b=1 , \cdots , r
                                \ \ \ \ \ n=0 , 1 , \cdots , \infty
\eeq
where $F^{ab}$ is the $N \times N$ matrix of elements
\beq
  (F^{ab})_{ij} = f_{ij}^{ab} = \xi_i^b \eta_j^a.
\eeq
We define the generating functional  of the currents $J_n^{ab}$. It is the
$r \times r$ matrix $T(z)$
of elements
\beq
  T^{ab}(z) = - \demi \delta_{ab} - \sum_{n \in \N} \frac{1}{z^{n+1}} J_n^{ab}
            = - \demi \delta_{ab} + tr \left ( \frac{1}{L - z} F^{ab} \right ).
\eeq
\proclaim Proposition.
On the manifolds $\Si$ we have the following two properties:
\begin{enumerate}
\item The currents $J_n^{ab}$ Poisson commute with all the quantities of
the form $tr(L^n).$
\item The generating functional $T(z)$ satisfies the defining relation of a
(classical) Yangian algebra:
\beq
  \label{Y}
  \{ T(y) \stackrel{\otimes}{,} T(z) \} = [ R(y,z) , T(y) \otimes T(z) ]
\eeq
with
\beq
  R(y,z) = - 2 \frac{\Pi}{y-z} \ , \ \ \ \ \ \Pi=\sum_{a,b=1}^r
                                                 e_{ab} \otimes e_{ba}.
\eeq
\end{enumerate}
\par

{\bf Proof.}
To prove this proposition we need the Poisson brackets
\bea
  \label{r1}
  \{ L_1 , L_2 \} & = & [r_{12}^0,L_1]-[r_{21}^0,L_2]
  + \sumij (f_{ii}-f_{jj}) \frac{1}{\sinh^2(q_{ij})} e_{ij} \otimes e_{ji} \\
  \label{r2}
  \{ L_1, F_2^{ab} \} & = & [-r_{21}^0+{\cal C}, F_2^{ab}] \\
  \{ F_1^{ab}, F_2^{cd} \} & = & (\delta_{ad} F_1^{cb} - \delta_{cb} F_2^{ad})
           \ {\cal C}.
\eea
Remark that the currents $J_n^{ab}$ and the Hamiltonians $tr(L^n)$ are
invariant under the symmetry
$$ \xi_i^a \longrightarrow \la_i \ \xi_i^a , \ \ \ \ \
   \eta_i^a \longrightarrow \la_i^{-1} \eta_i^a . $$
Therefore we can compute their Poisson brackets on the reduced phase
space straightforwardly; restricting ourselves to the manifolds
$f_{ii}=\alpha$, the last term in eq.(\ref{r1}) vanishes, and we will
systematically drop its contribution in intermediate calculations.

We emphasize that in eq.(\ref{r1},\ref{r2}) the same $r$-matrix appears.
Moreover it is the term $ [{\cal C}, F_2^{ab}]$ in eq.(\ref{r2})
which is responsible for the quadratic form of eq.(\ref{Y}), as we shall see
in what follows.

Introducing the generating functional $H(z)=tr(\frac{1}{L-z})$ of the
Hamiltonians $tr(L^n)$ we compute
\bean
  \{ \frac{1}{L_1-y} F_1^{ab} , \frac{1}{L_2-z} \} & = &
  - \left [ \frac{1}{L_2-z} \ r_{12}^0 \ \frac{1}{L_2-z} \ , \
            \frac{1}{L_1-y} F_1^{ab} \right ]
  + \left [ \frac{1}{L_1-y} \ r_{21}^0 \ \frac{1}{L_1-y} F_1^{ab} \ , \
            \frac{1}{L_2-z}  \right ]  \\
   & & + \frac{1}{L_1-y} \frac{1}{L_2-z} \ [{\cal C},F_1^{ab}] \
                       \frac{1}{L_2-z}.
\eean
Taking the trace we obtain
$$ \{ T^{ab}(y) , H(z) \} = tr \left ( F^{ab} \left [
                   \frac{1}{L-y} , \frac{1}{(L-z)^2} \right ]
                            \right ) = 0. $$
This proves the first part of the proposition. To prove the second part
we evaluate
\bean
  \{ \frac{1}{L_1-y} F_1^{ab} , \frac{1}{L_2-z} F_2^{cd} \} & = &
  - \left [ \frac{1}{L_2-z} \ r_{12}^0 \ \frac{1}{L_2-z} F_2^{cd} \ , \
            \frac{1}{L_1-y} F_1^{ab} \right ] \\
  & & + \left [ \frac{1}{L_1-y} \ r_{21}^0 \ \frac{1}{L_1-y} F_1^{ab} \ , \
            \frac{1}{L_2-z} F_2^{cd} \right ]  \\
  & & + \frac{1}{L_1-y} \frac{1}{L_2-z} \left ( \delta_{ad} F_1^{cb}
                     - \delta_{cb} F_2^{ad} \right ) \ {\cal C} \\
  & & + \frac{1}{L_1-y} \frac{1}{L_2-z} \left \{
         [{\cal C},F_1^{ab}] \ \frac{1}{L_2-z} F_2^{cd} \
       - \ [{\cal C},F_2^{cd}] \ \frac{1}{L_1-y} F_1^{ab} \right \}.
\eean
Hence taking the trace we get
\bean
  \{ T^{ab}(y),T^{cd}(z) \} & = & tr \left ( \frac{1}{L-y} \frac{1}{L-z}
    ( \delta_{ad} F^{cb} - \delta_{cb} F^{ad} ) \right ) \\
 & & + tr \left ( \frac{1}{L-y}
   \left [ \frac{1}{L-z} F^{cd} \frac{1}{L-z} \ , \ F^{ab} \right ] \right )
     - tr \left ( \frac{1}{L-z}
   \left [ \frac{1}{L-y} F^{ab} \frac{1}{L-y} \ , \ F^{cd} \right ] \right ).
\eean
Using the cyclicity of the trace and
$$ \frac{1}{L-y} \frac{1}{L-z} = \frac{1}{y-z}
   \left ( \frac{1}{L-y} - \frac{1}{L-z} \right ) $$
this becomes
\bean
  \{ T^{ab}(y),T^{cd}(z) \} & = & \frac{1}{y-z}
            \left ( \delta_{ad} ( T^{cb}(y) - T^{cb}(z) )
                   - \delta_{cb} ( T^{ad}(y) - T^{ad}(z) ) \right ) \\
  & & + \frac{2}{y-z} tr \left (
         \frac{1}{L-y} F^{cd} \frac{1}{L-z} F^{ab}
       - \frac{1}{L-y} F^{ab} \frac{1}{L-z} F^{cd}  \right ).
\eean
Remarking that
\bean
  tr \left ( \frac{1}{L-y} F^{cd} \frac{1}{L-z} F^{ab} \right ) & = &
   \sum_{ijkl=1}^N \left ( \frac{1}{L-y} \right )_{ij} \ \xi_j^d \
   \eta_k^c \ \left ( \frac{1}{L-z}\right )_{kl} \ \xi_l^b \ \eta_i^a \\
  & = & \left ( \sum_{ij=1}^N \left ( \frac{1}{L-y} \right )_{ij} \
   \xi_j^d \ \eta_i^a \right ) \left ( \sum_{kl=1}^N \left ( \frac{1}{L-z}
   \right )_{kl} \ \xi_l^b \ \eta_k^c \right ) \\
  & = &  \left ( T^{ad}(y) + \demi \delta_{ad} \right )
         \left ( T^{cb}(z) + \demi \delta_{cb} \right )
\eean
we prove the result (\ref{Y}).$\Box$

The rational limit is obtained by applying the canonical transformation
\bean
  p_i & \longrightarrow & \frac{1}{\epsilon} \ p_i \\
  q_i & \longrightarrow & \epsilon \ q_i
\eean
and sending $\epsilon$ to zero. In this limit
\bean
  L_0 & \longrightarrow & \frac{1}{\epsilon} \ L_{\rm rational} \\
  r_{12}^0 & \longrightarrow & \frac{1}{\epsilon} \ r_{12 \ {\rm rational}}^0.
\eean
The Casimir term drops therefore from eq.(\ref{r2}), leaving us with a
linear Poisson algebra
\beq
  \{ T(y) \stackrel{\otimes}{,} T(z) \}
   = - \demi [ R(y,z) , T(y) \otimes 1 + 1 \otimes T(z)]
\eeq
which is the result found by Gibbons and Hermsen.

\section{Conclusion}
The Euler-Calogero-Moser model is becoming more and more interesting.
On the one hand the computation of the classical $r$-matrix is made
considerably easier by the existence of the extra variables $f_{ij}$, the
more complicated $r$-matrix of the Calogero-Moser model following naturally
from a Hamiltonian reduction procedure. On the other hand, this model
exhibits an exact infinite symmetry which is just a current algebra
symmetry in the rational case and becomes an exact Yangian symmetry in
the trigonometric case. This structure is very much reminiscent of the one
discovered in \cite{BGHP}. Actually the two currents $J_0$ and $J_1$ (which
generate the full algebra) are identical in the two cases. Indeed, in our
case we have
$$ J_1^{ab} = \sumi p_i f_{ii}^{ab}
            - \sumij \frac{{\rm e}^{-q_{ij}}}{\sinh(q_{ij})} f_{ij}
              f_{ji}^{ab}. $$
Setting $ (X_i)^{ab} = f_{ii}^{ab}, \
          \Theta_{ij} = 2 \frac{{\rm e}^{2q_j}}{{\rm e}^{2q_i}
                                               -{\rm e}^{2q_j}} $
and using eq.(\ref{sm}) we can rewrite
$$ J_1^{ab} = \sumi p_i X_i^{ab} - \sumij \Theta_{ij} (X_i X_j)^{ab}. $$
This is exactly the current found in \cite{BGHP}. In fact the model
considered in \cite{MP,HW,BGHP} is a quantum version of our model for a
particular choice of orbit.

At this point two interesting problems arise. One is the understanding of
the role of the $r$-matrix in the quantization of these models. The other is
the hypothetical extension of these results to the elliptic case, which
still remains quite mysterious.

\bigskip

\bigskip

\noindent{\Large \bf Acknowledgements} $\ $ We thank D. Bernard for
discussions.

\bigskip

\bigskip

\noindent {\Large \bf Appendix}

\bigskip

\noindent The Weierstrass $\si$ function of periods
$ 2 \omega_1, 2 \omega_2 $ is the entire function defined by
\beq
  \si(z) = z \prod_{m,n\neq 0} \left ( 1 - \frac{z}{\omega_{mn}} \right )
  \exp \left [  \frac{z}{\omega_{mn}}
              + \demi \left ( \frac{z}{\omega_{mn}} \right ) ^2 \right ]
\eeq
with $ \omega_{mn} = 2 m \omega_1 + 2 n \omega_2 .$
The functions $\zeta$ and $\wp$ are
\beq
  \zeta(z) = \frac{\si'(z)}{\si(z)} \ , \ \ \ \ \ \wp(z) = -\zeta'(z),
\eeq
these functions having the symmetries
\beq
  \si(-z) = - \si(z)\ , \ \ \ \ \ \zeta(-z) = - \zeta(z) \ , \ \ \ \ \
  \wp(-z) = \wp(z).
\eeq
Their behaviour at the neighbourhood of zero is
\beq
  \si(z) = z + O(z^5) \ , \ \ \ \ \  \zeta(z) = z^{-1} + O(z^3) \ ,
  \ \ \ \ \ \wp(z) = z^{-2} + O(z^2).
\eeq
Setting
\beq
  l(q,\la) = - \frac{\si(q+\la)}{\si(q)\ \si(\la)}
\eeq
it is easy to check that
\beq
  \label{a1}
  l(-q,\la) = - l(q,-\la) \ , \ \ \ \ \
  l'(q,\la) = l(q,\la) \ [\zeta(\la+q) - \zeta(q)].
\eeq
We need several non trivial relations:
\beq
  - \frac{\si(\la-\mu) \ \si(\la+\mu)}{\si^2(\la) \ \si^2(\mu)} =
  \wp(\la) - \wp(\mu),
\eeq
\beq
  \label{a3}
  \si(x-y) \si(x+y) \si(z-t) \si(z+t)
  + \si(y-z) \si(y+z) \si(x-t) \si(x+t)
  + \si(z-x) \si(z+x) \si(y-t) \si(y+t) =0,
\eeq
\beq
  \frac{\si(2z) \ \si(x+y) \ \si(x-y)}{\si(x+z) \ \si(x-z) \ \si(y+z)
  \ \si(y-z)} = \zeta(x+z) - \zeta(x-z) + \zeta(y-z) - \zeta(y+z),
\eeq
this last equation becoming, in terms of the $l(q,\la)$ function,
\bea
  \label{a4}
  \frac{l(q,\la) \ l(-q,\la-\mu)}{l(q,\mu)} =
    \zeta(\la) + \zeta(\mu-\la) + \zeta(q) - \zeta(\mu+q).
\eea
Choosing the periods $\omega_1=\infty$ and $\omega_2=i \frac{\pi}{2}$ we
obtain the hyperbolic case
\beq
  \label{tr1}
  \si(z)= \sinh(z) \exp \left ( - \frac{z^2}{6} \right ) \ , \ \ \ \ \
  \zeta(z)=\coth(z) - \frac{z}{3} \ , \ \ \ \ \
  \wp(z)=\frac{1}{\sinh^2(z)} + \frac{1}{3}
\eeq
and
\beq
  \label{tr2}
  l(q,\la)=-\frac{\sinh(\la+q)}{\sinh(\la) \sinh(q)} \exp \left (
  - \frac{\la \ q}{3} \right ).
\eeq
All these formulas were collected in \cite{S2}.

\end{document}